\documentclass[12pt]{spieman}  
\usepackage{amsmath,amsfonts,amssymb}
\usepackage{graphicx}
\usepackage{setspace}
\usepackage{tocloft}
\usepackage[utf8]{inputenc}
\usepackage[left]{lineno}
\usepackage{hyperref}
\usepackage{cite}

\title{Curved holographic optical elements from a geometric view point: graphical visualization and parametric optimization}

\author[a,*]{Tobias Graf}
\author[b,c]{Dominik Hayd}
\affil[a]{Robert Bosch GmbH, Corporate Research and Advance Development, Robert-Bosch-Campus 1, 71272 Renningen, Germany}
\affil[b]{Robert Bosch GmbH, Technical Vocational Training Feuerbach, Stuttgart, Germany}
\affil[c]{Duale Hochschule Baden-Württemberg (DHBW) Stuttgart, Stuttgart, Germany}

\cftpagenumbersoff{figure}
\cftpagenumbersoff{table} 

\newtheorem{problem}{Problem}

\begin{document} 
\maketitle
\begin{abstract}
	We present numerical experiments regarding the diffraction characteristics and their precompensation arising from changing the macroscopic geometry of a flat HOE into a sphere segment. In particular, we discuss a parametric optimization scheme using a grating vector field model which was previously introduced in Ref. \citen{graf_curved_2021-1}.
\end{abstract}


{\noindent \footnotesize\textbf{*} Tobias Graf,  \linkable{tobias.graf3@de.bosch.com} }

\begin{spacing}{1}   


\section{Introduction}
Already in the 1970s holographic optical elements (HOE) have been employed in augmented reality (AR) displays, e.g. in head-up displays (HUD) for fighter jets \cite{close_holographic_1975}. While HUDs as well as head-worn or helmet mounted displays (HWD, HMD) have originated in military applications, they have since entered civilian domains in aerospace, automotive and more recently even in consumer products, prominent examples being Microsoft's HoloLens and MagicLeap's One devices as well as the North focals. In this paper, we will discuss applications of a geometric framework developed for modeling, simulation, and design of curved HOE, in particular with regard to curved holographic combiners for retina scanner displays (RSD) \cite{graf_curved_2021-1}.

\begin{figure}
	\centering\includegraphics[scale = 1.0]{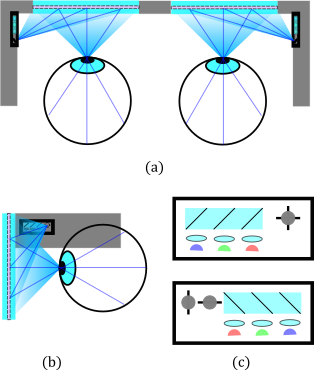} 
	\caption{The basic RSD architecture consists of a MEMS laser projection module and a HOE combiner in the lens to redirect the diffracted light towards the users eye. (a) top-view of a binocular RSD; (b) side view; (c) schematic illustration of MEMS based laser projetion modules using a single 2D scanning mirror or a combination of two 1D scanning mirrors in combination with laser diodes, collimating lenses and dichromatic reflectors for beam combining.} \label{fig:RSD}
\end{figure}

In a RSD \cite{viirre_virtual_1998, kollin_optical_1995}, a laser projection module comprised of laser sources, collimating, beam combining and beam shaping optics and MEMS (microelectromechanical systems) scanning mirrors is designed to scan the emitted laser light across the spectacle lenses containg a holographic combiner optics, e.g. a volume hologram, to redirect the scanned beams towards the user's pupil. The light passing the pupil creates a visual stimulus directly on the retina, which, using a high scanning frequency compared to the latency of the visual system, can create the impression of a virtual image for the user, cf. Fig. \ref{fig:RSD}.

\begin{figure}
	\centering
	\includegraphics[width = 0.8\textwidth]{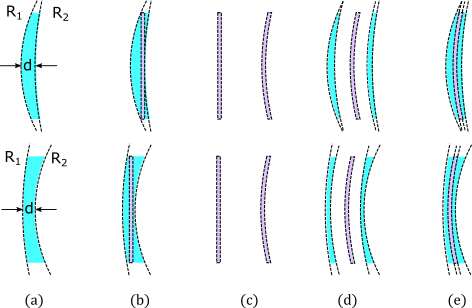}
	\caption{Curving a HOE before integration in a corrective lens can positively impact lens thickness and weight. (a) the power of a spherical lens is determined by the two outer radii, the thickness, and the refractive indices of the lens and the surrounding medium. (b) Embedding a flat HOE may require a thicker lens. (c) Deforming a flat HOE into a sphere segment after recording. (d) Integration of a curved HOE in a lens with curved mechanical interfaces matching the HOE curvature. (e) Curved HOE allow for integration in a thinner lens.} \label{fig:embeddingHOE}
\end{figure}

In recent years, new holographic materials, in particular photopolymers, have become commercially available \cite{jurbergs_new_2009} along with industrial processes for mass production of HOE \cite{bjelkhagen_mass_2017, hrabovsky_performance_2017} and their embedding in ophtalmic lenses \cite{korner_casting_2018}. In particular, there has been significant progress towards the industrialization of the integration of HOEs into ophtalmic lenses \cite{korner_casting_2018}. While this allows for vision correction in a RSD device, the process described in Ref. \citen{korner_casting_2018} typically requires a flat HOE to be curved after the recording process in order to be integrated in the corrective lens. Aside from technical aspects, curved HOE also offer a clear advantage over flat HOE with respect to the realizable lens thickness, which is directly related to aesthetics and weight, cf. Fig. \ref{fig:embeddingHOE}. Overall, the development of small form factor smartglasses leveraging the miniaturization potential of the RSD architectures has progressed significantly in recent years.  

The paper is organized as follows:
\begin{itemize}
	\item In Section \ref{sec:sciVis} we introduce a visualization tool developed in Python to demonstrate the implementation of the methods introduced previously\cite{graf_curved_2021-1}. We restate the forward and the inverse problem for curved HOE formulated in Ref. \citen{graf_curved_2021-1} and visualize them graphically using our tool.
	\item In Section \ref{sec:paramOpt} we turn to the question of finding a recording configuration for a flat HOE to achieve optimal diffraction characteristics after curving the element. We describe a parametric optimization approach using the geometric methods \cite{graf_curved_2021-1} based on grating vector fields.
	\item In Section \ref{sec:sciVisUseCase} we reevaluate selected examples obtained during the parametric optimization described in Section\ref{sec:paramOpt} using our vsiualization tool. 
	\item We conclude our presentation with a discussion and conclusion in Section \ref{sec:conclusion} in particular with respect to open challenges and future research regarding the simulation and design of curved HOE and as well as visualization and engineering tools. 
\end{itemize}

\section{Scientific Visualization in Python} \label{sec:sciVis}
For our investigations regarding modelling and design approaches for curved HOE, we have implemented a geometric modelling approach based on a grating vector field formalism \cite{graf_curved_2021-1} in different variations. One variant consists of a Python based visualization tool which we will introduce here briefly to visualize the concepts behind the forward and the inverse problem stated in Ref. \citen{graf_curved_2021-1}, which we recall here briefly for completeness.

\subsection{Recap: Forward and Inverse Problem for Curved HOE} \label{ssec:fwinvpblm}
\begin{figure}
	\centering
	\includegraphics[scale = 0.8]{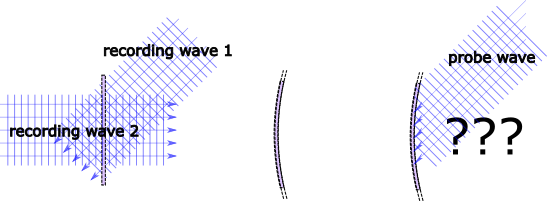} 
	\caption{Forward problem of HOE deformation: After recording an HOE with two coherent recording waves (often referred to as reference and object waves in the holography literature), the macroscopic geometry is changed. The forward problem consists of the task to predict the diffraction of a probe wave at the deformed HOE.} \label{fig:prblm-fw}
\end{figure}

We recall the so-called forward problem for curved HOE, which was previously introduced \cite{graf_curved_2021-1}. An illustration of the forward problem for curved HOE is shown in Figure \ref{fig:prblm-fw}.

\begin{problem} \label{pblm:fw} 
	Given a planar HOE $\mathcal{H}$ and a prescribed in advance surface $\mathcal{S}$, what are the optical characteristics of the HOE $\tilde{\mathcal{H}}$ obtained after deforming $\mathcal{H}$ into the shape $\mathcal{S}$?     	
\end{problem}

\begin{figure}
	\centering
	\includegraphics[scale = 0.8]{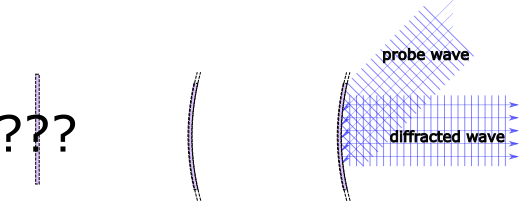}
	\caption{Inverse problem of HOE deformation: Given a prescribed in advance macroscopic recording and reconstruction geometry as well as a prescribed in advance diffraction behavior of the HOE during reconstruction (i.e. its micrsocopic structure), the inverse problem consists of the task to determine the microscopic geometry of the HOE in the recording geometry prior to deformation into the reconstruction geometry. } \label{fig:prblm-inv}
\end{figure}

We also recall the associated inverse problem, which was also introduced in Ref. \citen{graf_curved_2021-1}. Figure \ref{fig:prblm-inv} graphically illustrates the inverse problem.

\begin{problem} \label{pblm:inv} 
	Given a surface $\mathcal{S}$, prescribed in advance probe wave $\mathbf{E}_\textrm{p}$  (or familiy of reference wavefronts, e. g. for polychromatic systems) in combination with a prescribed in advance target wavefront $\mathbf{E}_\textrm{d}$ (or family of target wavefronts), find an HOE $\mathcal{H}$ such that the curved HOE resulting from reshaping  $\mathcal{H}$ into the geometry of $\mathcal{S}$ results in a HOE which transforms $\mathbf{E}_\textrm{p}$ into $\mathbf{E}_\textrm{d}$.	 
\end{problem}

\subsection{Visualization of Flat and curved HOE}
\begin{figure}
	\centering
	\includegraphics[width= \textwidth]{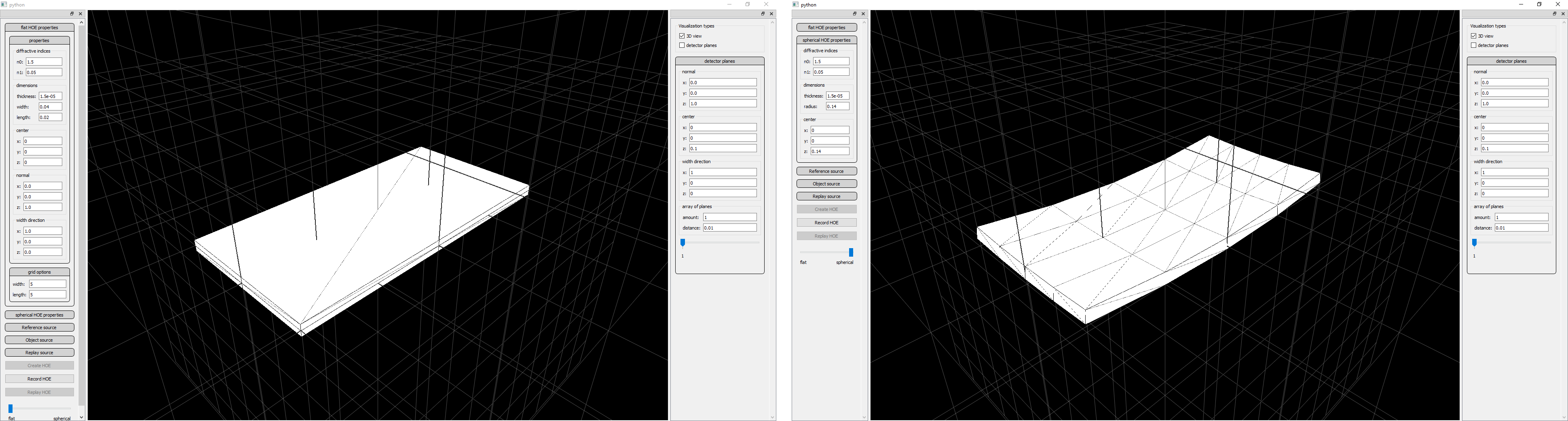}
	\caption{Left: flat HOE geometry, Right: curved HOE geometry} \label{fig:flat-and-curved-HOE}
\end{figure}

\begin{figure}
	\centering
	\includegraphics[width=1.0\textwidth]{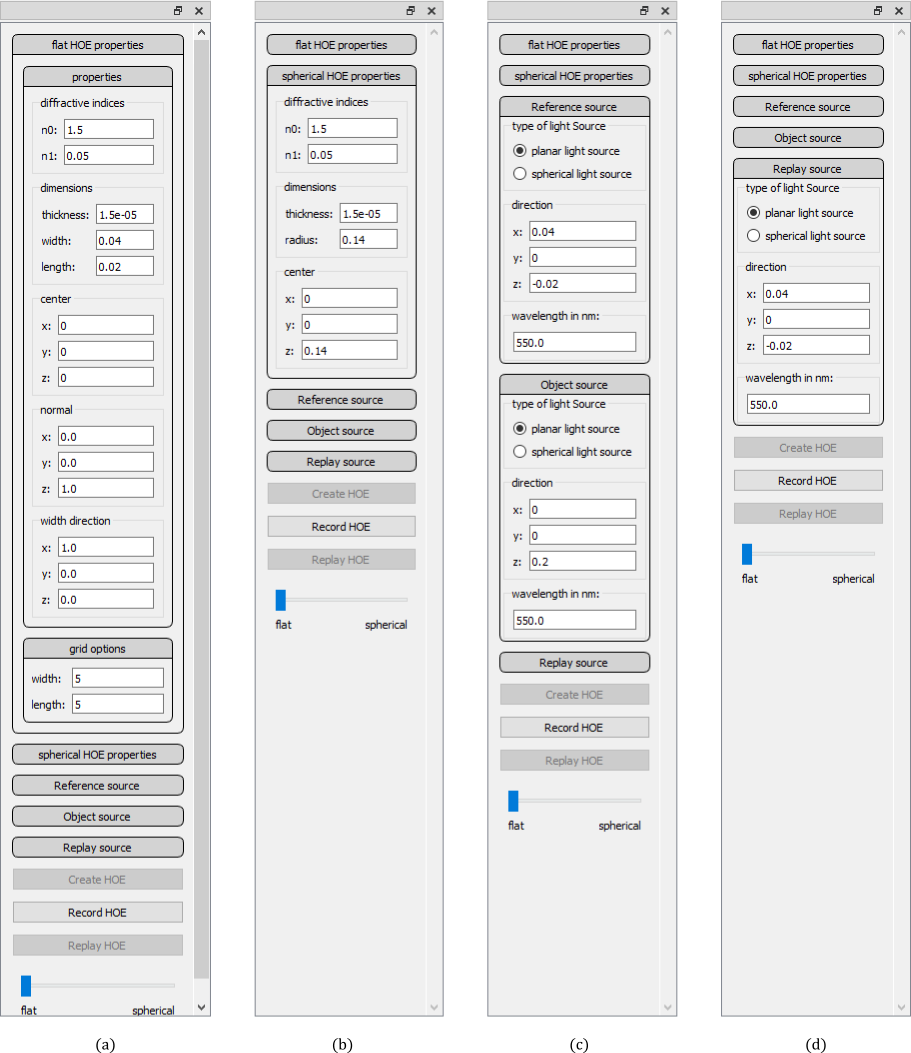}
	\caption{Property tabs from left to right: (a) flat HOE geometry and sampling; (b) curved HOE geometry; (c) recording wavefront settings; (d) probe wavefront settings} \label{fig:proptabs}
\end{figure}

As shown in the two screenshots in Figure \ref{fig:flat-and-curved-HOE}, the tool comprises a graphical user interface (GUI) which contains various properties tabs and simulation buttons on the left, a graphical visualization display in the center, and a detector and visualization settings tab on the right. The basic HOE geometry including the sampling is defined in the \emph{flat HOE properties} tab shown in Fig. \ref{fig:flat-and-curved-HOE} and in a close up view in Figure \ref{fig:proptabs} (a). The geometry of the curved HOE is currently limited to projecting the flat HOE surface onto a sphere segment along the surface normal direction of the flat HOE plane. The geometry of the sphere containing the curved HOE is defined in the \emph{spherical HOE properties} tab visible in the screenshot on the right-hand side in Fig. \ref{fig:flat-and-curved-HOE} and in close-up in Fig. \ref{fig:proptabs} (b). The geometry can be changed from flat to curved or from curved to flat interactively using a slider contained in the properties tab. Depending on the current state of the program, different options for user actions are accessible via buttons. These are
\begin{itemize}
	\item \emph{Create HOE}: to generate the initial geometry of a HOE or update the geometry after adjusting parameters;
	\item \emph{Record HOE}: Computing the grating vector field for the current geometry settings, in particular including the choice of a curved or flat HOE, either intitially or after changing the settings in one or both of the recording waves.
	\item \emph{Replay HOE}: Compute the diffracted wave vectors based on the current grating vector field and the current probe wave settings.
\end{itemize} 

\subsection{Recording and deforming a flat HOE}
\begin{figure}
	\centering
	\includegraphics[width= \textwidth]{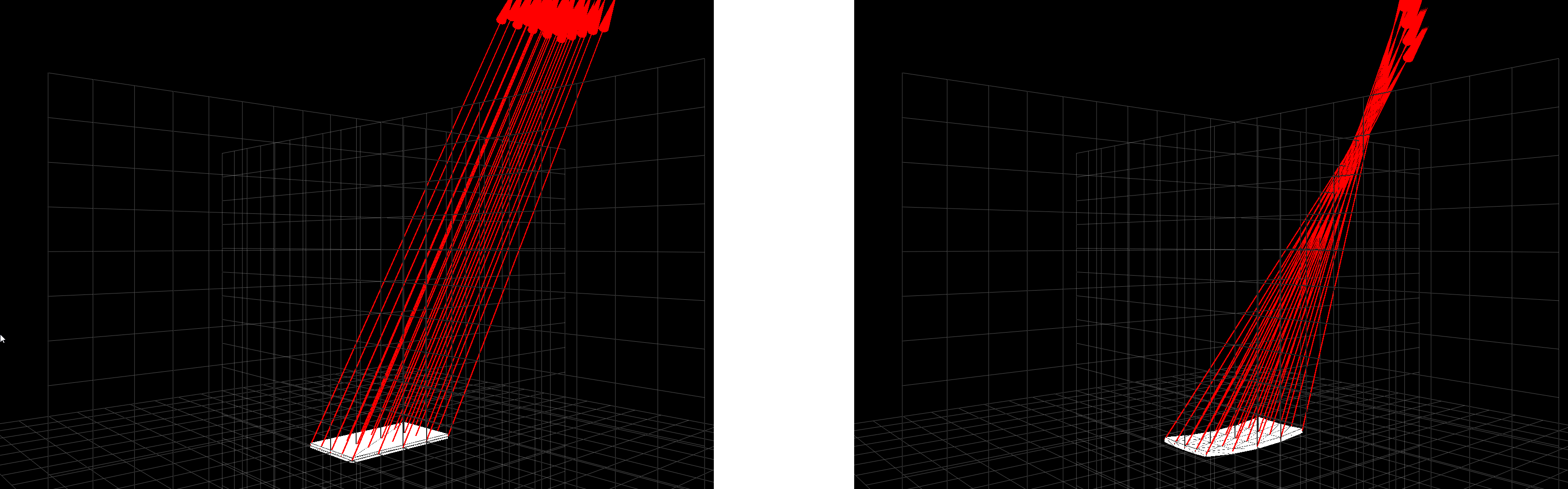}
	\caption{Left: the grating vector field obtained from recording waves in a flat HOE; Right: the perturbed grating vector field for an HOE which was deformed inta a curved shape after recording} \label{fig:flat_vs_curved_HOE_GratingVF}
\end{figure}

We return now to the forward problem which we have restated above as Problem \ref{pblm:fw}. As an exmaple, we consider a volume grating obtained from the interference of two plane waves, one propagating in direction $(0.04, 0.0, -0.02)$ and the other one propagating in direction $(0.0, 0.0, 0.2)$. The grating vector field $\mathbf{k}_\textrm{g}$ is computed from the recording wave vectors as
\begin{equation}
\mathbf{k}_\textrm{g} = \mathbf{k}_\textrm{obj} -  \mathbf{k}_\textrm{ref},
\end{equation}
see Ref. \citen{graf_curved_2021-1} and references therein for more details on coupled wave theories in HOE modelling. The properties tab for the recording waves, called reference and object wave in the GUI, are shown in Fig. \ref{fig:proptabs} (c) for the case of plane wave illumination. Fig. \ref{fig:flat_vs_curved_HOE_GratingVF} shows on the left the grating vector field of the flat HOE after recording. Note that the grating vectors all appear to be colinear as is to be expected from a volume grating. In the screenshot on the right-hand side of Fig. \ref{fig:flat_vs_curved_HOE_GratingVF} we see the perturbed grating vector field after curving the flat HOE. As stated already above, the grating vector field deformation is computed based on orthogonal moving frames \cite{graf_curved_2021-1}. The change in the macroscopic geometry of the HOE surface results in a relative reorientation of the grating vectors describing the local grating geometry.

\begin{figure}
	\centering
	\includegraphics[width= \textwidth]{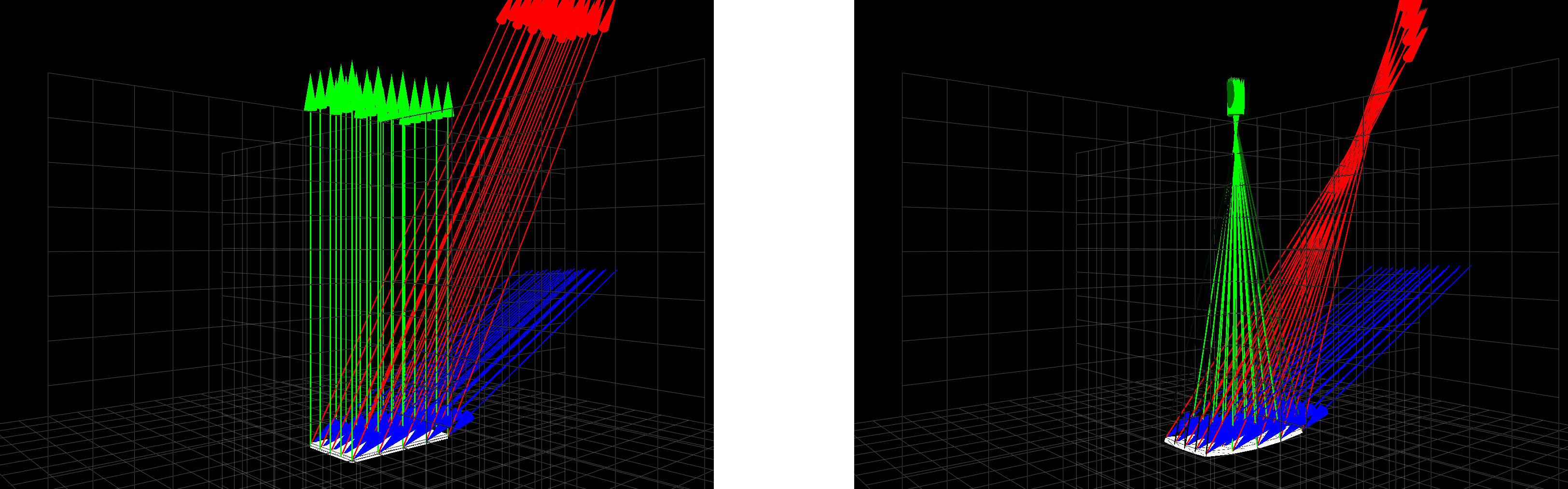}
	\caption{Left: the grating vector field (red) obtained from recording waves in a flat HOE along with the probe wave vectors (blue), and the diffracted wave vectors (green); Right: the perturbed grating vector field (red) for an HOE which was deformed inta a curved shape after recording along with the probe wave vectors (blue), and the diffracted wave vectors (green)} \label{fig:lat_vs_curved_HOE_Diffraction}
\end{figure}

Once we have created the HOE, i.e. computed the grating vector field, we can reconstruct the diffracted wave from a probe wave. The settings for the probe wave (or replay wave) are defined in the \emph{Replay Source} properties tab shown in Fig. \ref{fig:proptabs} (d). If we use the same settings for the probe wave as for one of the recording waves (e.g. the reference wave), we replay the flat HOE in on-Bragg condition, assuming that any fabrication tolerances like shrinkage etc. can be neglected. Fig. \ref{fig:lat_vs_curved_HOE_Diffraction} shows on the left the replay of the flat HOE and on the right the replay of the HOE curved after recording the grating vector field. Note that in the flat HOE case the colinear probe wave vectors are diffracted into colinear wave vectors as is to be expected. For the curved HOE, we observe that the diffracted wave vectors are no longer colinear. Moreover, as is indicated by the different color tones, the Bragg condition is violated resulting in locally varying diffraction efficiencies indicated by darker wave vectors in the diffracted vector field.  

The grating vector field on the curved HOE is computed based on a one-to-one correspondence between orthogonal frames \cite{graf_curved_2021-1}. In other words, we use cylindrical coordinates
\begin{align}
x & = r \cos(\theta), \\
y & = r \sin(\theta), \\
z & = z
\end{align}
and consider on the flat HOE at each point the local frame $\{ \mathbf{t}_r, \mathbf{t}_\theta, \mathbf{n} \}$
\begin{equation} \label{eq:orthoframe-flat}
\begin{split}
\mathbf{t}_r & = (\sin(\theta), \cos(\theta), 0), \\
\mathbf{t}_\theta & = (\cos(\theta), -\sin(\theta), 0), \\
\mathbf{n} & =  (0, 0, 1).
\end{split}
\end{equation}
The curved HOE with radius of curvature $R$ can be parametrized in cylindrical coordinates e.g. as
\begin{equation}
f(r, \theta) = R - \sqrt{R^2 - (r*sin(\theta)^2) - (r*cos(\theta)^2)}.
\end{equation}
We consider on the curved HOE the associated frame $\{\tilde{\mathbf{t}}_r, \tilde{\mathbf{t}}_\theta, \tilde{\mathbf{n}} \}$
\begin{equation} \label{eq:orthoframe-curved}
\begin{split}
\tilde{\mathbf{t}}_r & = \frac{(\sin(\theta), \cos(\theta), f_{r}(r, \theta))}{\sqrt{1 + (f_r(r, \theta))}}, \\
\tilde{\mathbf{t}}_\theta & = \frac{(\cos(\theta), -\sin(\theta), f_{\theta}(r, \theta))}{\sqrt{1 + (f_\theta(r, \theta))}}, \\
\tilde{\mathbf{n}} & = \tilde{\mathbf{t}}_r \times \tilde{\mathbf{t}}_\theta
\end{split}
\end{equation}
The grating vector 
\begin{equation} \label{eq:gvf-flat}
\mathbf{k}_g(r, \theta) = g_1(r, \theta)\mathbf{t}_r(r, \theta) + g_2(r, \theta) \mathbf{t}_\theta(r, \theta) + g_3(r, \theta) \mathbf{n}(r, \theta)
\end{equation}
on the flat HOE is transformed into the grating vector
\begin{equation} \label{eq:gvf-curved}
\tilde{\mathbf{k}}_g(r, \theta) = g_1(r, \theta) \tilde{\mathbf{t}}_r(r, \theta) + g_2(r, \theta) \tilde{\mathbf{t}}_\theta(r, \theta) + g_3(r, \theta) \tilde{\mathbf{n}}(r, \theta)
\end{equation}
on the curved HOE.

\subsection{Recording and flattening of a curved HOE}
\begin{figure}
	\centering
	\includegraphics[width= \textwidth]{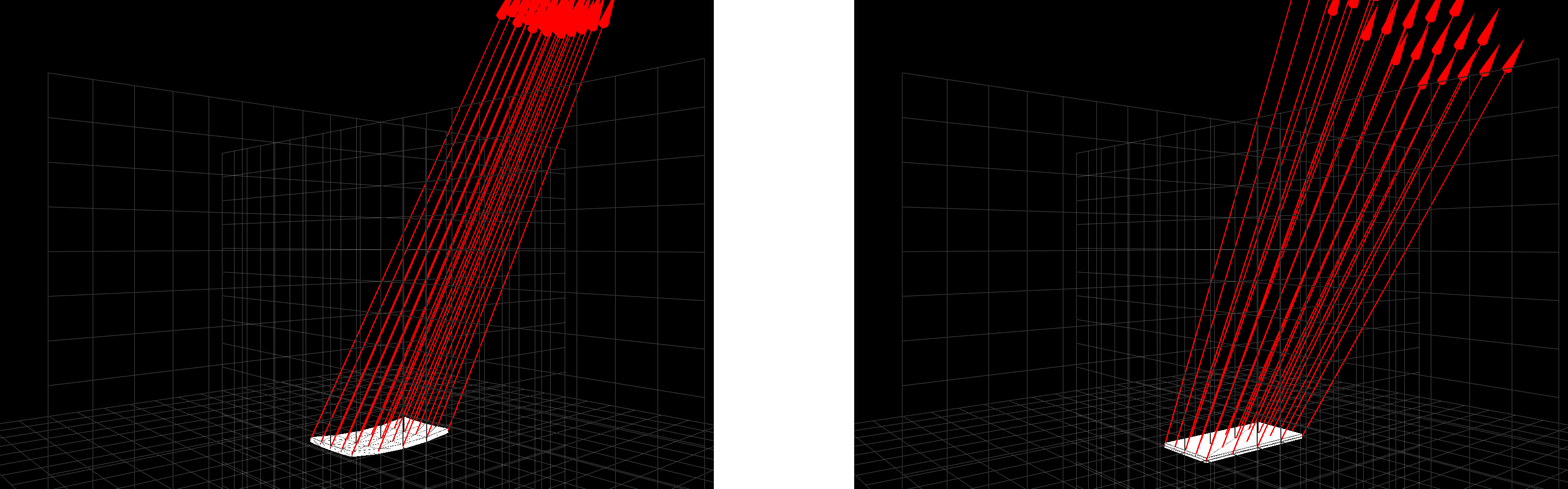}
	\caption{Left: the grating vector field obtained from planar recording waves in a curved HOE; Right: the perturbed grating vector field for an HOE which was obtained by flattening the curved HOE after recording.} \label{fig:curved_vs_flattened_HOE_GratingVF}
\end{figure}

We shift now our attention to the inverse problem restated above as Problem \ref{pblm:inv}. Thus we start with a curved HOE geometry and record the grating vector field obtained from the recording wave vectors. As is shown in Fig. \ref{fig:curved_vs_flattened_HOE_GratingVF} on the left, the grating vectors are all colinear in the case of a volume grating recorded with two planewaves in the recording setup. Once this HOE is flattened, the grating vector field is perturbed as is visible in Fig. \ref{fig:curved_vs_flattened_HOE_GratingVF} on the right.

\begin{figure}
	\centering
	\includegraphics[width= \textwidth]{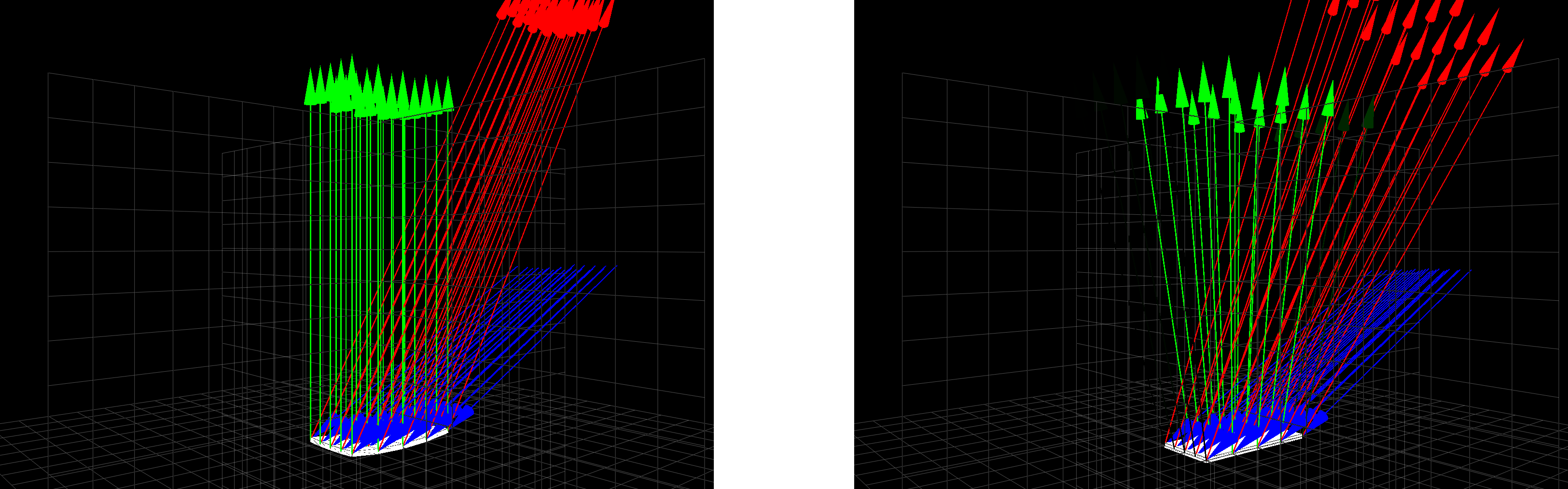}
	\caption{Left: the grating vector field (red) obtained from planar recording waves in a curved HOE along with the probe wave vectors (blue), and the diffracted wave vectors (green); Right: the perturbed grating vector field for an HOE which was obtained by flattening the curved HOE after recording  along with the probe wave vectors (blue), and the diffracted wave vectors (green)} \label{fig:curved_vs_flattened_HOE_Diffraction}
\end{figure}

The effects of replaying the (ideal) curved HOE and the associated flattened HOE are shown in Fig. \ref{fig:curved_vs_flattened_HOE_Diffraction}. The more interesting question in many design situations, however, is how to realize the flattened grating vector field in a recording setting. A surprisingly practical approach based on parametric optimization using forward calculations is discussed therefore in the following section.

Note that in this case, the grating vector field during the recording step is expressed in the curved frame , cf \eqref{eq:orthoframe-curved} and then transformed during the flattening step using the frame in \eqref{eq:orthoframe-flat}. In other words, the grating vector field of the curved HOE is presented in the form of \eqref{eq:gvf-curved} and then transformed into the flattened grating vector field in the form of \eqref{eq:gvf-flat}.	

\section{Parametric Optimization in Matlab} \label{sec:paramOpt}
We consider the inverse problem for an HOE which is expected to transform a divergent spherical wave centered at a point $(x_\textrm{probe}, y_\textrm{probe}, z_\textrm{probe})$ into a diffracted wave which is ideally a convergent spherical wave converging towards the point $(x_\textrm{diffr}, y_\textrm{diffr}, z_\textrm{diffr})$. For simplicity, we restrict the discussion to the monochromatic case at $639$ nm wavelength. A sample choice of coordinates which is suitable e.g. for lab demonstrator setups of RSDs is contained in Table \ref{tab:paramopt-target}.

\begin{table}
	\centering
	\begin{tabular}{c | c | c}
		$(x_\textrm{probe}, y_\textrm{probe}, z_\textrm{probe})$ & $(x_\textrm{diffr}, y_\textrm{diffr}, z_\textrm{diffr})$ & Radius of curvature (HOE) \\
		\hline 
		$(-44.0, 0.0, 41.0)$ & $(0.0, 0.0, 25.0)$ & $140$
	\end{tabular}
	\caption{Point source location for probe wave and diffracted wave with respect to HOE center in millimeter (mm). The wavelength is held constant at $639$ nm. The probe wave is divergent, the target wavefront for the diffracted wave is a convergent spherical wave.} \label{tab:paramopt-target}
\end{table}

The optical design challenge consists in determining a recording configuration for a flat HOE which meets the above diffraction characteristics, based on an existing recording setup which interferes two spherical waves, one divergent and one convergent, with centers located at $(x_\textrm{ref}, y_\textrm{ref}, z_\textrm{ref})$ and $(x_\textrm{obj}, y_\textrm{obj}, z_\textrm{obj})$, respectively. The baseline coordinates for the recording setup are summarized in Table \ref{tab:paramopt-baselineRecordingSetup}.

\begin{table}
	\centering
	\begin{tabular}{c | c| c | c}
		Configuration & $(x_\textrm{ref}, y_\textrm{ref}, z_\textrm{ref})$ & $(x_\textrm{obj}, y_\textrm{obj}, z_\textrm{obj})$ & $(x_\textrm{probe}, y_\textrm{probe}, z_\textrm{probe})$ \\
		\hline 
		Baseline & $(-44.0, 0.0, 41.0)$ & $(0.0, 0.0, 30.0)$ & $(-44.0, 0.0, 41.0)$ \\
	\end{tabular}
	\caption{Point source location for reference, object, and probe waves with respect to HOE center in millimeter (mm). The wavelength is held constant at $639$ nm. Reference and probe wave are divergent, the object wave is convergent.} \label{tab:paramopt-baselineRecordingSetup}
\end{table}

As we have shown in Section \ref{sec:sciVis}, we could compute the grating vector field for a flattened HOE using geometric methods. However, this result does not give us immediate information on how to realize such a flattened HOE with a recording setup. Therefore, we present here an approach based on parametric optimization which was implemented in a Matlab script. 

The underlying idea of this parametric optimization approach is to allow for severeal parameter variations in the recording setup. For the present discussion we consider variations in $x$ and $z$ coordinate of the refernce source and in the $z$ coordinate of the object source by $\pm 10$ mm about the baseline configuration, cf. Table \ref{tab:paramopt}. We compute the diffraction of the reference wave for a family of curved HOE obtained after deforming flat HOE recorded with these parameter variations using the by now familiar geometric framework.

\begin{table}
	\centering
	\begin{tabular}{c | c| c | c }
		Coordinate & Baseline & Perturbation & \# of steps  \\
		\hline 
		$x_\textrm{ref}$ & $-44.0$ & $\pm 10$ & 5  \\
		$y_\textrm{ref}$ & $0.0$ & --- & ---   \\
		$z_\textrm{ref}$ & $41.0$ & $\pm 10$ & 5   \\
		$x_\textrm{obj}$ & $0.0$ & --- &  ---  \\
		$y_\textrm{obj}$ & $0.0$ & --- & ---  \\
		$z_\textrm{obj}$ & $30.0$ & $\pm 10$ & 5 
	\end{tabular}
	\caption{Point source location for reference, object, and probe waves with respect to HOE center in millimeter (mm). The wavelength is held constant at $639$ nm. Reference and probe wave are divergent, the object wave is convergent.} \label{tab:paramopt}
\end{table}

In order to decide on a promising candidate for manufacturing, and with regard to automatization using numerical optimization schemes, we will discuss next a set of merit criteria which we expect to be promising candidates for numerical optimization schemes in the optical design of curved HOE.

\begin{figure}
	\centering
	\includegraphics[width= \textwidth]{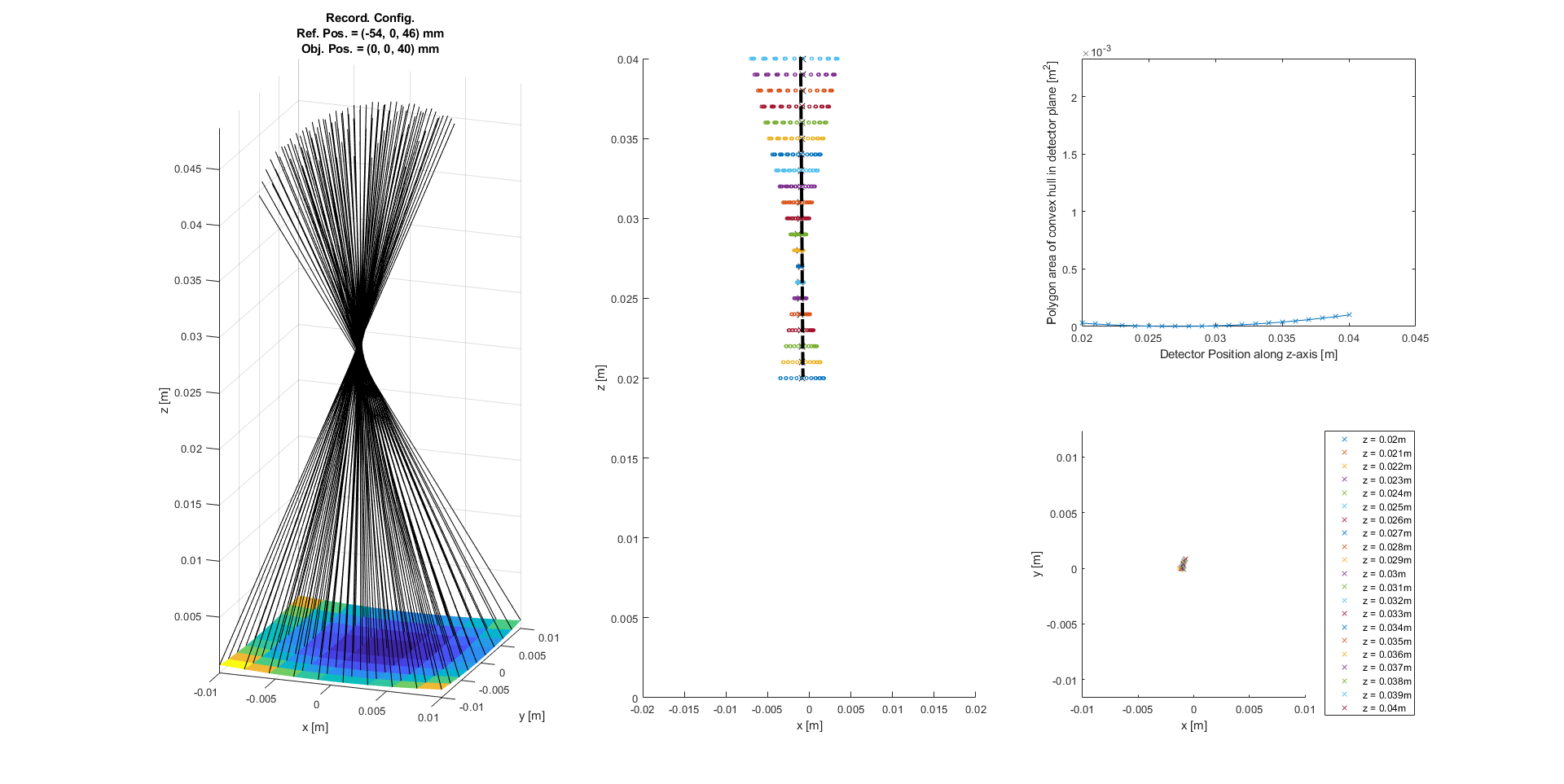}
	\caption{Analysis of the \emph{best choice} candidate for a recording configuration to precompensate for deformation of the HOE material after recording. Left: HOE surface and direction of propagation of the diffracted wave vectors; Center: intersection points of diffracted rays with a familiy of parallel detector planes and approximate axis through the bary centers in each detector plane; Right: polygon area of the convex hull of the ray interception points in the detector panes (top); and xy-coordinates of the barycenters of the ray interception points in the detector planes.} \label{fig:paramopt-hoe-config-20}
\end{figure}

\begin{figure}
	\centering
	\includegraphics[width= \textwidth]{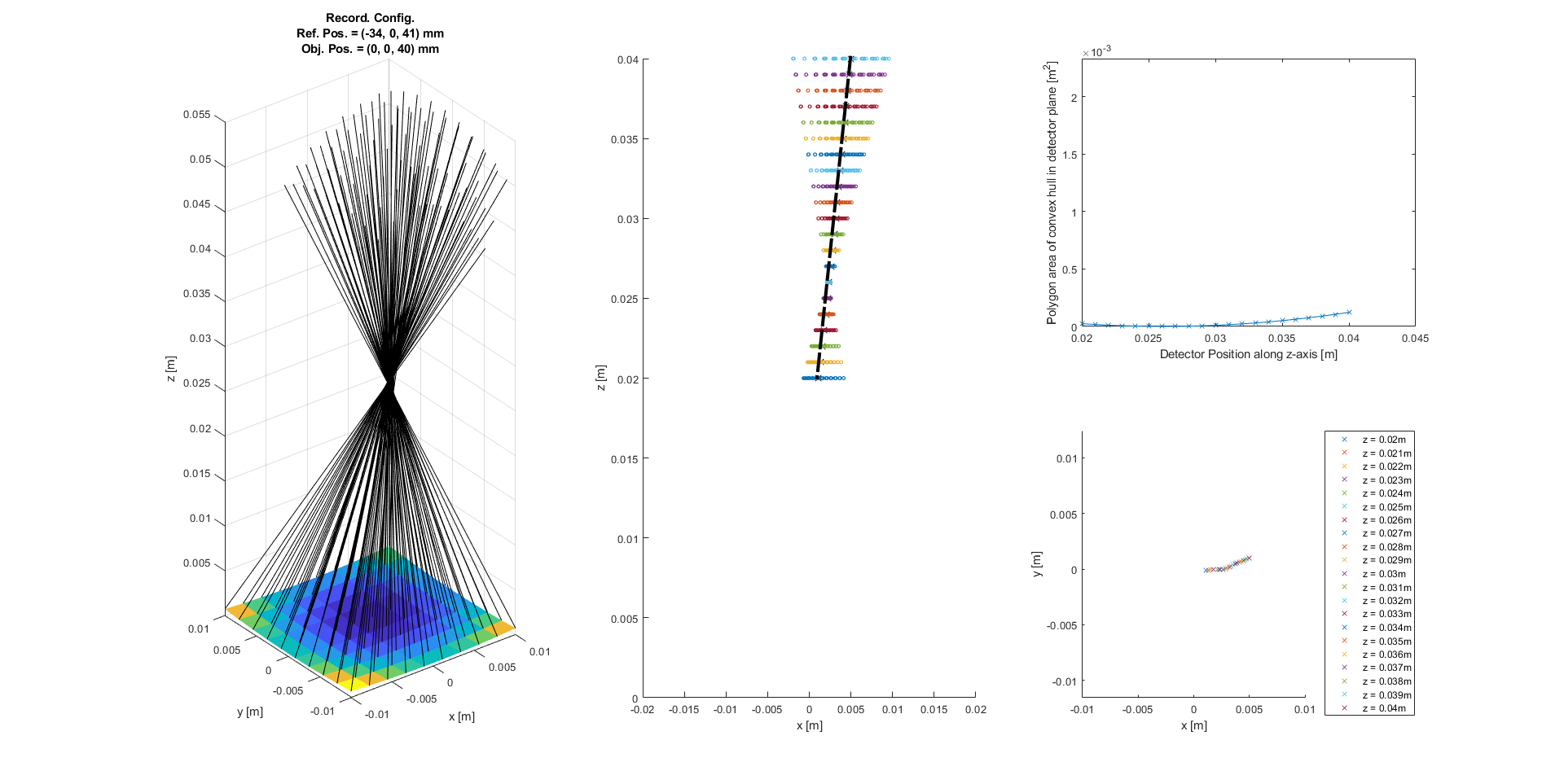}
	\caption{Analysis of a \emph{dismissed} candidate for a recording configuration to precompensate for deformation of the HOE material after recording. Left: HOE surface and direction of propagation of the diffracted wave vectors; Center: intersection points of diffracted rays with a familiy of parallel detector planes and approximate axis through the bary centers in each detector plane; Right: polygon area of the convex hull of the ray interception points in the detector panes (top); and xy-coordinates of the barycenters of the ray interception points in the detector planes.} \label{fig:paramopt-hoe-config-115}
\end{figure}

Although we expect that the geometric framework under discussion can be modified to allow for the analysis of wavefront aberrations for HOE \cite{peng_imaging_1983, peng_nonparaxial_1986, verboven_aberration_1986}, we focus in the present investigation on ray-based merit criteria, where the rays are dervived from the diffracted wave vectors. In particular, we use a family of detector planes perpendicular to the surface normal of the HOE in the center and evaluate the ray interception points of the rays with these detector planes. The area of the convex hull of the interception points is regarded as a measure of how strongly the light is focused in this detector plane. The barycenters of the convex hulls in the detector planes are regarded as a measure for the skewness of the cone formed by the diffracted rays. In Fig. \ref{fig:paramopt-hoe-config-20} and Fig. \ref{fig:paramopt-hoe-config-115}, the graphics on the right show the curved HOE surface and the diffracted ray cone, the graphics in the center show the ray interception points in the different detector planes and the bary centers. The approximate axial orientation of the barycenter is indicated with a dashed black line. The graphics on the right-hand side show at the top the polygon area of the convex hulls of the ray interception points in each detector plane. The graphics at the bottom right show the $(x,y)$ coordinates of the barycenters in the detector planes.  

As is observable from the different visualizations and the merit function plots, the diffraction behavior in Fig. \ref{fig:paramopt-hoe-config-20} is significantly closer to the target configuration of a spherical wave, i.e. a cone of rays, converging towards the point $(0,.0, 0.0, 25.0)$ mm with respect to the HOE center. Moreover, in order to move away from a graphical analysis, the merit criteria can in principle be combined into a scalar valued merit function to shift to a numerical optimization scheme. While the polygon area is already scalar valued, the suitable way to transfer the barycenter coordinates into a scalar value is given by computing the distance of the barycenters to a reference axis (e.g. the $z$-axis is the present situation).  

\section{Use Case: Interactive Analysis of Matlab Results} \label{sec:sciVisUseCase}
We briefly return to the anaylsis capabilities of the visualization tool, and investigate here the two curved HOE obtained from the different recording configurations already discussed in the context of our parametric optimization approach in Section \ref{sec:paramOpt}. We have constructed the two flat HOE corresponding to the different recording configurations and replayed them with the same probe wave (i.e. the reference wave) after introducing a curvature radius of $140$ mm. The wave property settings are summarized in Table \ref{tab:overviewConfigs}.

\begin{table}
	\centering
	\begin{tabular}{c | c| c | c}
		Configuration & $(x_\textrm{ref}, y_\textrm{ref}, z_\textrm{ref})$ & $(x_\textrm{obj}, y_\textrm{obj}, z_\textrm{obj})$ & $(x_\textrm{probe}, y_\textrm{probe}, z_\textrm{probe})$ \\
		\hline 
		Baseline & $(-44.0, 0.0, 41.0)$ & $(0.0, 0.0, 30.0)$ & $(-44.0, 0.0, 41.0)$ \\
		Precompensated & $(-54.0, 0.0, 46.0)$ & $(0.0, 0.0, 40.0)$ & $(-44.0, 0.0, 41.0)$ \\
		Offender & $(-34.0, 0.0, 41.0)$& $(0.0, 0.0, 40.0)$& $(-44.0, 0.0, 41.0)$ 
	\end{tabular}
	\caption{Point source location for reference, object, and probe waves with respect to HOE center in mm. The wavelength is held constant at $639$ nm. Reference and probe wave are divergent, the object wave is convergent.} \label{tab:overviewConfigs}
\end{table}

\begin{figure}
	\centering
	\includegraphics[width= \textwidth]{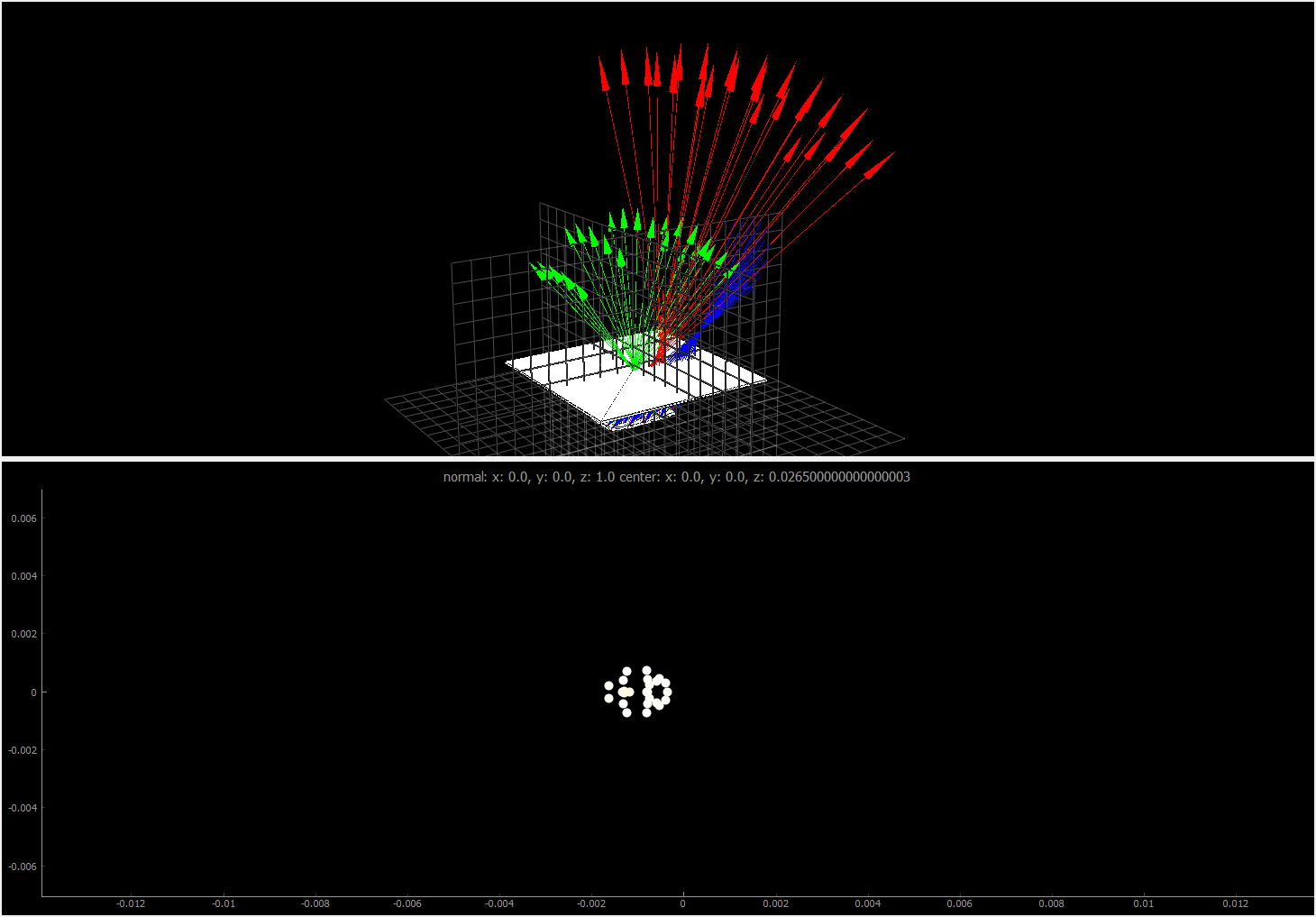}
	\caption{Grating vector field, probe wave vector field, diffracted wave vector field (top) and interception points of the diffracted wave vector directions in a detector plane (bottom) for the \emph{best choice} precompensation candidate.} \label{fig:precom-geom-diffr-detetor}
\end{figure}

Fig. \ref{fig:precom-geom-diffr-detetor} shows the HOE geometry, grating and wave vector fields and a detector plane of the \emph{best choice} candidate determined with Matlab. The Detector plane view suggest a narrowing of the diffracted wave vector directions at around 26.5 mm along $z$, however, the graphical visualization sucggests that a perfect focus is not achieved in accordance with our expectation from the Matlab computations.

\begin{figure}
	\centering
	\includegraphics[width= \textwidth]{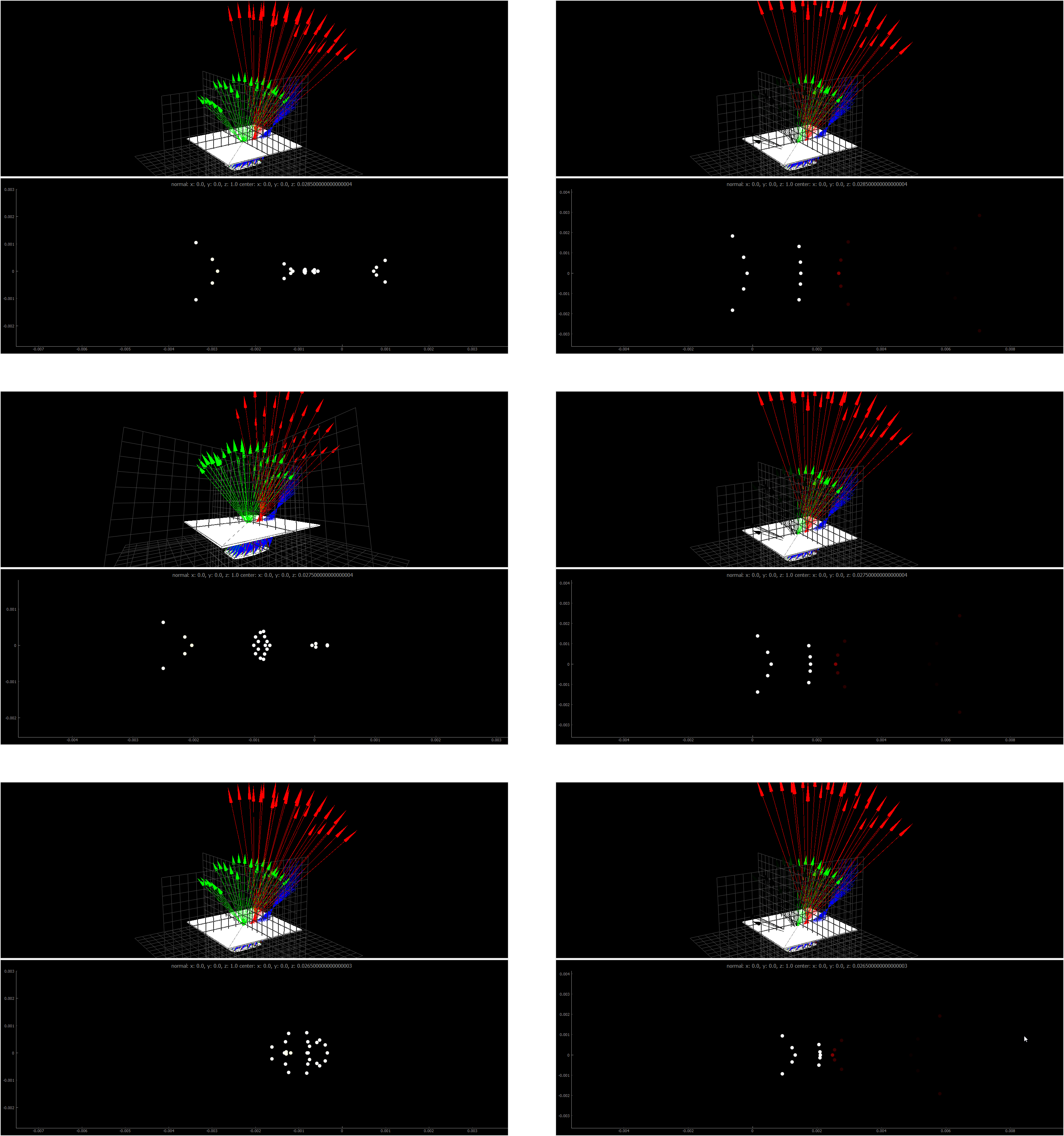}
	\caption{Interception points of the diffracted wave vector directions with three different detector planes for 
		Left: \emph{best choice} and Right: \emph{dismissed} precompensation candidate. The interception points are color coded according to the diffraction efficiency at the HOE ranging from black ($0.0$) through red and yellow to white ($1.0$).} \label{fig:compare-detectorviews} 
\end{figure}

A more detailed look for three different $z$-heights of the detector planes are shown in Fig \ref{fig:compare-detectorviews} for the \emph{best choice} candidate on the left and for the \emph{dismissed} candidate on the right. The results in Fig \ref{fig:compare-detectorviews} confirm the hypothesis that the parameter variation in the recording setup can be used to adjust for the focusing behavior of the diffracted wave if the flat HOE is curved after recording.

Moreover, the 3D visualizations and detector plane views clearly show the potential benefit of including the diffraction efficiency into the merit function. For example, the variation of diffraction efficiencies appears to more uniform across the diffracted directions for the \emph{best choice} candidate than for the \emph{dismissed} candidate, cf. Fig. \ref{fig:compare-detectorviews}. We point out here, that the Matlab based approach presented in Section \ref{sec:paramOpt} did not account for diffraction efficiencies. It seems rather appropriate though, to use the diffraction efficiences as weights for the ray intercept points when computing merit functions -- such as barycenters -- in future investigations. In contrast the detector planes in  of Fig. \ref{fig:compare-detectorviews} show a much less uniform efficiency distribution and the patterns formed by the ray interception points seem indicative of more complex wave front aberrations. 

We want to point out here, that the deformation model used in the Matlab script differed from the one used in Section \ref{sec:sciVis}. In fact, the deformed grating vector field on the curved HOE in the current section was computed using directional derivatives in in Cartesian coordinates $(x,y)$ not in cylindrical coordinates as discussed in Section \ref{sec:sciVis}. This approach leads to a set of non-orthogonal frames for the curved HOE
\begin{equation}
\begin{split}
\tilde{\mathbf{t}}_x & = \frac{(1, 0, f_x(x,y))}{\sqrt{1 + (f_x(x,y))^2}}, \\
\tilde{\mathbf{t}}_y & = \frac{(0, 1, f_y(x,y))}{\sqrt{1 + (f_y(x,y))^2}}, \\
\tilde{\mathbf{n}} & = \mathbf{n} 
\end{split}
\end{equation}
where $\tilde{\mathbf{n}}$ is again the surface normal as in \eqref{eq:orthoframe-curved}. Since the frame $\{ \tilde{\mathbf{t}}_x, \tilde{\mathbf{t}}_y, \tilde{\mathbf{n}} \}$ is not orthogonal, the transformation from the flat to the curved grating vector field analogous to the transition from equation \eqref{eq:gvf-flat} to equation \eqref{eq:gvf-curved}, will not be length preserving in general. However, as long as the the radius of curvature $R$ of the curved HOE is large enough compared to the lateral dimensions of the HOE, the angle betweeen the two tangent vector $\tilde{\mathbf{t}}_x $ and $\tilde{\mathbf{t}}_y $ differs only slightly from a $\frac{\pi}{2}$. We will see in Section \ref{sec:sciVisUseCase} that the precompensated configuration found with the nonorthoganl model seems to mantain its characteristic properties also in an analysis with the orthogonal frames. This is to be expected, since the deformation of planar HOE foil in to a smooth spherical segment must inherently result in some local deformation, such as stretching, due to Gauss's \emph{theorema egregium} \cite{carmo_differential_2018}. Our numerical obersavations thus correspond to the experimental observations that functional curved HOE can be obtained from flat ones, if the radius of curvature is comparatively large compared to the lateral HOE dimensions.

\section{Conclusion} \label{sec:conclusion}
In the present investigation, we have evaluated different aspects related to the implementation and application of geometric methods for the simulation and design of curved HOE. We are convinced that towards the goal of a highly automated design and optimization process for curved HOE, there is a need to develop and implement mathematical methods and numerical algorithms for physical simulations and numerical optimizations on the one hand, but also a need to support optical engineers with adequate and easy to use visualization tools for non-standard analyses and custom design studies.

We believe that the work presented here can serve as a motivation for applied mathematicians to develop adequate mathematical models, a priori error estimates, and numerical algorithms, as well as for optical designers and engineers to think about new ways and methods to model, compute, and design HOEs in particular as transparent combiner elements in AR displays. 

Among the potential topics for further improvements we plan to investigate in the future alternative diffraction theories for volume holograms as well as a validation of the efficiency calculations.  Another open point is currently the handling of the zero order light in the visualization and analysis. The inclusion of zero order is light is important for example  in system analyses regarding stray light. However, this issue might be solved by integrating the new geometric framework into an existing software tool which is capable of handling multiple light paths and ray splitting. Several aspects of potential improvements of the merit functions used in the HOE design process and the possibility to move towards more rigorous numerical optimization schemes have already been hinted at in the above discussion.

As is clear from the discussion in Sections \ref{sec:paramOpt} and \ref{sec:sciVisUseCase}, the geometric approach discussed here is expected to be compatible with ray-tracing as well as field tracing methods. In particular the possibility to combine a computationally efficient HOE model with state-of-the-art multiconfiguration optimization and physical optics beam propagation methods is expected to provide powerful tools to support optical engineers in their tasks. We are thus convinced, that computational geometry and discrete differential geometry will prove themselves once again as useful toolboxes for the development of new simulation and design methods for curved HOE. 

\section*{Acknowledgments}

T.~G. would like to thank Christian Hellmann of LightTrans UG, Reinhold Fieß and Stefanie Hartmann at Bosch Corporate Research, Tobias Werner at Bosch Car Multimedia as well as Johannes Hofmann at Bosch Sensortec for interesting discussions on computational optics and holography.

D.~H. would like to thank Prof. Janko Dietzsch of DHBW Stuttgart and Florian Wolff of Bosch technical vocational training for making the work on this topic possible. In addition, thanks go to Hendrik Jansen, Lorenz Jobst and Christopher Narr for discussions and support during the development of the visualization tool.




\end{spacing}
\end{document}